\documentstyle[12pt]{article}

\def\br{\begin{thebibliography}{abc}}
\def\er{\end{thebibliography}}

\begin{document}

\thispagestyle{empty}
\setcounter{page}{0}
\bibliographystyle{unsrt}
\footskip 1.0cm
\thispagestyle{empty}
\setcounter{page}{0}
\begin{flushright}
SU-4240-615\\
July, 1995 \\
\end{flushright}
\centerline {\bf A DISCRETIZED VERSION OF KALUZA-KLEIN THEORY }
\centerline {\bf WITH TORSION AND MASSIVE FIELDS    }
\vspace*{10mm}
\centerline {\bf Nguyen Ai Viet \footnote{ On leave of absence from the High
Energy Division, Centre of Theoretical Physics, P.O.Box 429 Bo Ho 10000, Hanoi
Vietnam.}
\footnote{ Address after September 1, 1995: Physics Department, Rockerfeller
University, New York, NY, USA }
and Kameshwar C.Wali }
\vspace*{3mm}
\centerline {\it  Department of Physics, Syracuse University,}
\centerline {\it Syracuse, NY 13244-1130, U.S.A.}
\vspace*{10mm}
\normalsize
\centerline {\bf Abstract}
We consider an internal space of two discrete points in the fifth dimension of
the Kaluza-Klein theory by using the formalism of noncommutative geometry
developed in a previous
paper \cite{VIWA} of a spacetime supplemented by two discrete points. With the
nonvanishing internal torsion 2-form there are no constraints implied on the
vielbeins. The theory contains a pair of tensor, a pair of vector and a pair
of scalar fields. Using the generalized Cartan structure equation we are able
not
only to determine uniquely the hermitian and metric compatible connection
1-forms, but also the nonvanishing internal torsion
2-form in terms of vielbeins. The resulting action has a rich and complex
structure, a particular feature being the existence of massive modes. Thus the
nonvanishing internal torsion generates a Kaluza-Klein type model with zero and
massive modes.

\bigskip

\newpage

\section{ Introduction}
It is generally believed that our current description of spacetime underlying
both classical physics as well as quantum field theories is
unsatisfactory and inadequate to deal with the description of phenomena at
short distances. One is seeking a mathematical formalism that provides a
quantum
description of natural phenomena that, a priori, does not speak about spacetime
in its basic formulation, but spacetime of classical physics as well as quantum
field theories emerges in certain limiting regimes- just as classical behaviour
of quantum systems can emerge in certain limiting regimes \cite{QUAN}. The
recent
proposal of Connes \cite{CO} and the so called noncommutative geometry
(NCG) appears very promising towards the achievement of such a goal. It has
given rise to the description of the Standard Model \cite{COLO} with new
insights as regards spontaneous symmetry breaking and quark and lepton masses.
It is natural to ask whether and how the classical general relativity fits into
the scheme of NCG.

The first step in answering this question was taken by Chamseddine et al
\cite{CFF}, whose starting point was an abstract two-sheeted continuum that
could be considered as the direct product of a single spacetime continuum and
two discrete
points. This led to gravity coupled to a Brans-Dicke scalar field. The
scalar field can be interpreted as the distance between the two sheets
\footnote{ More recently, other authors using different approaches have
obtained
essentially the same result. See Ref.\cite{VIWA} for references to related work
.}.

Similarly, it is always extremely tempting to give geometrical meaning to other
physical fields.
Thus, in the traditional Kaluza-Klein
theory massless tensor, vector and scalar fields together with their massive
excitations
appear as result of extending the physical four-dimensional spacetime by an
additional continuous fifth dimension.
Unfortunately, the massive
modes are infinite in number. In a previous paper \cite{VIWA},
we have developed the formalism for a discretized version of Kaluza-Klein
theory within the framework of NCG. The starting point, as in Ref.\cite{CFF},
is an extended spacetime
that includes two discrete points of the continuous internal fifth dimension of
the Kaluza-Klein theory. We presented a generalization of the usual Riemannian
geometry in the new context that demanded a vielbein consisting, to begin with,
a pair of tensor, a pair of vector and a pair of scalar fields. Following the
usual steps in building a theory of gravitation with the new geometry, we
imposed torsion free, metric compatibility conditions on the connection 1-forms
from which we constructed the action through the Ricci scalar curvature.
We found that the imposed conditions altered the field content of the theory in
a dramatic way, requiring in addition to the tensor, vector and scalar fields,
new
dilaton-like dynamical fields. The connection 1-forms and hence the Ricci
scalar curvature were determined uniquely in terms of these fields. The
resulting action provided a rich structure that lent itself to intriguing
interpretations. One of the dilaton fields, for instance, could give rise to
masses and cosmological constant. Moreover by imposing a reality condition on
the vielbein 1-forms we could make the dilaton fields disappear leading to
the zero-mode sector of the Kaluza-Klein theory as in Ref.\cite{LVW}. The
previous NCG models that contain gravity coupled to
the Brans-Dicke scalar can be considered as a particular case when the vector
field is set to zero.

While these interpretations are interesting in
themselves to merit further study, we seek in this paper a formulation that
does not alter the initial field content of the theory of two independent
tensor, vector and scalar fields. From the viewpoint of the underlying
mathematical framework of NCG, this is a reasonable requirement: the vielbein
1-forms should be free of constraints, retaining their most general form.
The problem is how to achieve this. Now for physical reasons, it is necessary
that we
impose the metric compatibility condition. We recall that in the ordinary
Cartan-Riemannian geometry the vanishing of torsion yields unique connection
1-forms in terms of the metric coefficients and their derivatives.
Non-vanishing torsion requires additional
information besides the metric. In our formulation, we find a way to avoid this
situation. We impose a reality condition on the
connection 1-forms and release the strict torsion free condition.
In order to keep as close as possible to the usual Riemannian
geometry, we assume that the usual spacetime indexed torsion 2-forms do vanish.
However, we do not assume that the
discrete internal space indexed torsion 2-form vanishes.
This results in the unique determination of the related connection 1-forms.
As we shall see, the nonvanishing internal torsion 2-form can be also
determined in terms of the given vielbeins. This way we have an action that
describes the
general field content that we started with initially. The most remarkable
result is that this discrete version of Kaluza-Klein theory contains a finite
number of massive modes.

The paper is organized as follows: In the next section we will review briefly
the basic formalism and give the necessary formulas in order to make this paper
self-contained. In Sect.3, we discuss how we compute the connection 1-forms,
internal torsion and the Ricci scalar curvature. In Sect.4, we present the
general structure of the action and consider special cases. The final section
is devoted to a summary and conclusions.

\setcounter{equation}{0}
\section{ Two-point internal space and vielbein}
\subsection{ Algebra of smooth functions and generalized derivatives}
We consider a physical space-time manifold
${\cal M}$ extended by a discrete internal space of two points to which we
assign a $Z_2$-algebraic structure.
With this extended space-time, the customary algebra of smooth functions
 ${\cal C}^\infty ({\cal M})$ is generalized to ${\cal A} ~=~
{\cal C}^\infty({\cal M})\oplus{\cal C}^\infty({\cal M})$ and any
generalized function $F \in {\cal A}$ can be written as
\begin{equation}
F(x)= f_+(x) e + f_-(x) r~ ,
\end{equation}
where
$ e,r \in Z_2 = \{ e,r~|~ e^2=e~,~ r^2=e~,~ er=re=r ~\}$.
We adopt a $ 2 \times 2$ matrix representation for $ e,r $:
\begin{equation}
e =\pmatrix{1&0\cr
                0&1\cr}  ~~ ,~~ r=\pmatrix{1&0\cr
                                                 0&-1\cr}~.
\end{equation}
Then the function $ F(x)$ assumes a $2 \times 2$ matrix form,
\begin{equation}
F  ~=~f_+(x)\pmatrix{1&0\cr
       0&1\cr} ~ + ~ f_-(x) \pmatrix{1&0\cr
                                   0&-1\cr} ~=~\pmatrix{f_1(x)&0\cr
                                                0&f_2(x)\cr}~,
\end{equation}
where
\begin{equation}
 f_\pm(x)= 1/2.( f_1(x) \pm f_2(x) ) .
\end{equation}
In this paper we will use small
letters to denote the quantities of ordinary geometry and capital letters
for generalized quantities of NCG.

With the algebra ${\cal A}$ of smooth functions, we have what we may consider
as the algebra of the generalized 0-forms $ \Omega^0({\cal M})=
{\cal C}^\infty({\cal M})
\oplus {\cal C}^\infty({\cal M})$. To build the corresponding generalized
higher forms, we need an exterior derivative or the Dirac operator
$ D$ \cite{CO,COLO} in the language of NCG. For this purpose, as in
Ref.\cite{VIWA}, let us define derivatives $ D_N ( N= \mu, 5) $ by
\begin{eqnarray}\label{CODER}
&& D_\mu = \pmatrix{\partial_\mu &0\cr
                          0 &\partial_\mu\cr} ,~~~\mu = 0,1,2,3~,
\cr
&& D_5 = \pmatrix{0& m\cr
-m&0\cr}  ,
\end{eqnarray}
where $ m $ is a parameter with dimension of mass.
We specify the action of the derivatives on the 0-form elements as given by
\begin{equation}\label{COMDER}
D_N(F)= [D_N, F] ~~,~~N=\mu, 5 ~~~~,
\end{equation}
satisfying the Newton-Leibnitz rule,
\begin{equation}
D_N(FG) = D_N(F) G + F D_N(G).
\end{equation}

Then the exterior derivative operator $D$ is given by
\begin{eqnarray}
D \doteq (~DX^\mu D_\mu~+~ DX^5 \sigma^\dagger D_5~),
\end{eqnarray}
where
\begin{equation}
\sigma^\dagger ~=~ \pmatrix{ ~0 & -1\cr
                    1 & 0 \cr}.
\end{equation}
$ DX^M $ are in general $ 2 \times 2 $ matrices that form a basis of
the generalized 1-forms. They are direct generalizations of the
differential elements. When spacetime becomes curved, as in general
relativity (GR), $ DX^M $ denote a generalized curvi-linear differential
elements. Their
concrete form can be given in a concrete basis. The explicit form of $DX^M$ in
the orthonormal basis will be given in the next subsection.
\subsection{ General and orthonormal basis of 1-forms }
The possible metric structure
is guaranteed by the existence of a local orthonormal basis: the vielbein
$ E^A $. Analogously to GR, if we work in the locally flat basis the vielbein
$E^A$ can be chosen to be orthonormal. In Ref.\cite{VIWA}, we chose a diagonal
representation for the curvi-linear basis $DX^\mu$ and $DX^5\sigma^\dagger$ to
construct generalized one- and higher forms in analogy with the usual
Riemannian geometry. However, it is more convenient to work in a representation
in which the vielbeins $ E^A ( A= a,\dot 5) $ are diagonal. Locally, $E^A$ is
given as follows
\begin{eqnarray}
E^a &~~=~~& \pmatrix{e^a & 0 \cr
                     0  & e^a \cr} ~~,\cr
E^{\dot 5} &~~=~~& \pmatrix{ 0 & \theta \cr
                             \theta & 0 \cr}~~,
\end{eqnarray}
where $ e^a $ is some ordinary vierbein 1-forms and $\theta $ is some hermitian
Clifford element\footnote{
Completely, in analogy with GR, we can represent $ e^a $ and $ \theta $ as the
locally flat Dirac matrices $\gamma^a $ and $ \gamma^5 $ as in the spinorial
representation of Connes-Lott model.( This representation is used widely in
literature. See for example \cite{CO,COLO,CFF} for details. However, in our
formalism the two sheets are not necessarily the ones of different chiralities.
Hence $ \theta $ in general will be kept as an abstract Clifford element).}.

In this basis
the wedge product can be defined as follows
\begin{equation}\label{wedge}
E^A \wedge E^B = - E^B \wedge E^A.
\end{equation}
In the orthonormal and locally flat basis $E^A$, the curvi-linear differential
elements $ DX^M $ are in general not diagonal any more. Conversely, we can
choose to
work in the representation in which $ DX^M$ are diagonal. Then $E^A$ is not
diagonal anymore
as discussed in Ref.\cite{VIWA}. Both basis span the space of generalized
1-forms; hence
an arbitrary 1-form $ U $ in NCG is given by
\begin{equation}
U~~=~~E^A U_A ~~= DX^M U_M ,
\end{equation}
where $ U_A $ and $ U_M $ are the components of the 1-form $ U $
in the $E^A$ and $ DX^M $ basis respectively.
As $E^A$ and $DX^M$ themselves are also 1-forms, we can express them in terms
of each other as follows
\begin{eqnarray}
E^A~ &~~=~~& DX^M E^A_{~M} ~~,\cr
DX^M &~~=~~& E^A E^M_{~A} ~~,
\end{eqnarray}
where $ E^A_M $ and $E^M_A$ are generalized functions satisfying
\begin{eqnarray}
E^A_{~N} E^N_{~B} &~~=~~& \delta ^A_{~B} \cr
E^A_{~N} E^M_{~A} &~~=~~& \delta^M_{~N}~~  .
\end{eqnarray}
Without any loss of generality we can choose $ E^A_{~M} $ as follows :
\begin{eqnarray}\label{VIELBEIN}
E^a_\mu ~=~ \pmatrix{ e^a_{1\mu}(x) & 0 \cr
                              0 & e^a_{2\mu}(x) \cr}&~~,~~& E^a_5 ~~=~~0 \cr
E^{\dot 5}_\mu ~~=~~ \pmatrix{ a_{1\mu}(x) & 0 \cr
                                  0 & a_{2\mu}(x) \cr} = A_\mu &~~,~~&
E^{\dot 5}_5 = \pmatrix{ \varphi_1(x) & 0 \cr
                             0 & \varphi_2(x) \cr} ~~=~~\Phi,
\end{eqnarray}
( We use a ${\dot 5}$ index in the orthonormal basis to distinguish it from the
index $5$ in the curvi-linear basis ).
Thus
\begin{equation}\label{TRANS}
DX^\mu ~=~E^a E^\mu_a ~~,~~ DX^5~=(E^{\dot 5} - E^a A_a) \Phi^{-1},
\end{equation}
where $ A_a = E^\mu_a A_\mu $.

Now we can derive the transformation rules for the components of an arbitrary
1-form $ U $ between the two basis
\begin{eqnarray}
U_a ~~=~~ E^\mu_a ( U_\mu + A_\mu U_5) &~~~,~~~& U_{\dot 5} = \Phi^{-1} U_5 ~~,
\cr
U_\mu ~~=~~ E^a_\mu U_a - A_\mu \Phi U_{\dot 5} &~~~,~~~& U_5 =
\Phi U_{\dot 5}~~.
\end{eqnarray}

To this end we note that the exterior derivative of a general 1-form $ U
=DX^M U_M = E^A U_A $ is given by
\begin{eqnarray}
D U &~~=~~& (DX^\mu + DX^5\sigma^\dagger D_5 ) U ~~,\cr
&~~=~~& E^a\wedge E^b ~( D U)_{ab} + E^a\wedge E^{\dot 5}~ 2 ({\cal
D}U)_{a{\dot5}} ~~.
\end{eqnarray}

Using Eq.(\ref{TRANS}), we find
\begin{eqnarray}\label{DER1F}
( DU)_{bc} &~~=~~& {1\over 2} E^\mu_{~b} E^\nu_{~c}( \partial_\mu
E^a_{~\nu} - \partial_\nu E^a_{\mu}) U_a
- {1\over 2} E^\mu_{~b} E^\nu_{~c}
(\partial_\mu A_\nu -\partial_\nu A_\mu ) \Phi U_{\dot 5} \cr
&~~~+~&
{1\over 2} (E^\mu_{~b}  \partial_\mu U_c - E^\nu_{~c} \partial_\nu U_b)
+ {m \over 2} [ (A_b {\tilde E}^\nu_{~c} -A_c {\tilde E}^\nu_{~b})
E^a_{~\nu} U_a + ( A_c {\tilde U}_b - A_b {\tilde U}_c) \cr
&~~~+~& (A_b {\tilde E}^\nu_{~c} -A_c {\tilde E}^\nu_{~b})
({\tilde A}_\nu -A_\nu) \Phi U_{\dot 5}] , \cr
( DU)_{b{\dot 5}} &~~=~~& {1\over 2}~{\tilde E}^\mu_{~b}~
({\partial_\mu \Phi
\over\Phi} U_{\dot 5} +\partial_\mu U_{\dot 5})
+ {m\over 2}( \Phi^{-1}({\tilde U}_b -{\tilde E}^\mu_{~b}E^c_{~\mu}U_c) \cr
&~~~+~& {\tilde E}^\mu_{~b}\Big( A_\mu - {\tilde A}_\mu) (1+ {\tilde
\Phi}^{-1}\Phi)~\Big)U_{\dot 5} + {\tilde A}_b {\tilde U}_{\dot 5})~~ ,
\end{eqnarray}
where we have redefined $ A_\mu $ in Eq.(\ref{VIELBEIN}) as $-A_\mu\Phi^{-1} $.

In the $ E^A $ basis the hermitian conjugate of an arbitrary 1-form
$ U = E^a U_a + E^{\dot 5}U_{\dot 5} $ is the 1-form $ U^\dagger = E^a U_a +
E^{\dot 5} {\tilde U}_{\dot 5} $ where
\begin{equation}
{\tilde F} ~=~ \pmatrix{ f_2 & 0 \cr
                        0 & f_1\cr} ~~, ~~{\rm for~any~ function~}
F ~=~\pmatrix{f_1 & 0 \cr
               0 & f_2 \cr}~~.
\end{equation}

In the orthonormal basis, we have chosen $ E^A$ to be hermitian, the general
1-forms need not be hermitian, neither does the $DX^M$ basis.

\subsection{Generalized metric}
Following Ref.\cite{VIWA,LVW}, we define the metric ${\cal G}$ as the
sesquilinear inner product of two 1-forms $ U $ and $ V$ satisfying
\begin{eqnarray} \label{METRIC1}
< U~ F~,~ V~ G > & ~=~ & F < U~,~V> G ~~ , \cr
< U \otimes R ~,~ V\otimes S> &~=~& R^\dagger < U~,~V> S ,
\end{eqnarray}
where $F, G$ are functions and $R, S $ are 1-forms.
Assuming the existence of the local orthonormal basis $ E^A $, we have
\begin{equation} \label{METRIC2}
< E^A~, ~ E^B > ~~=~~ \eta^{AB},
\end{equation}
where $\eta^{AB} ~=~ signature(~-~,~+~+~+~+~) $.

{}From Eqs.(\ref{METRIC1}) and (\ref{METRIC2}) we obtain the generalized metric
tensor in the familiar form
\begin{eqnarray}
{\cal G }_{M N } &~=~& E^{A}_{~M} \eta_{AB} E^B_{~N}, \cr
{\cal G }^{M N}  &~=~& E^M_{~A} \eta^{AB} E^N_{~B}.
\end{eqnarray}
With the vielbeins given in Eq.(\ref{VIELBEIN}), the components of the metric
tensors ${\cal G}^{MN}$ and ${\cal G}_{MN}$ turn out to be
\begin{eqnarray}\label{METRIC}
{\cal G}^{\mu\nu} &~= ~& G^{\mu\nu} ~{\dot =}~  \pmatrix{ g_1^{\mu\nu} & 0 \cr
                          0 & g_2^{\mu\nu} \cr} , \cr
{\cal G}^{\mu 5} &~ =~ &  A^\mu ~=~ {\cal G}^{5 \mu} ~~  , \cr
{\cal G}^{55} & = & \Phi^{-2}+ A^2 ~~,\cr
{\cal G}_{\mu\nu} &~=~ &G_{\mu \nu} + A_\mu A_\nu ~{\dot =}~
\pmatrix{ g_{1\mu\nu} & 0 \cr
         0 & g_{2\mu\nu}\cr} +  A_\mu A_\nu ~~, \cr
{\cal G}_{\mu 5} &~ =~ &  {\cal G}_{5\mu} ~=~ A_{\mu}\Phi ~~,\cr
G_{55} & = &  \Phi^2  .
\end{eqnarray}
where $g_i^{\mu \nu} = e^\mu_{ia}\eta^{ab} e^\nu_{ib}~,~ i=1,2$ are
the metric tensors on the two sheets.

In passing we note that the components of the metric tensor are identical in
form with
those in the 5-dimensional Kaluza-Klein theory except that in the present case,
the usual continuous $x^5$-dependence is replaced by the matrix form.

\setcounter{equation}{0}
\section{ Connection, torsion and curvature}
\subsection {Hermitian and metric compatible connection 1-forms}
We have shown in Ref.\cite{VIWA,LVW} that the metric compatible or Levi-Civita
connection 1-form $\Omega_{AB}$ satisfies \footnote{ The interested reader
might see Ref.\cite{VIWA} for the relation between the connection and the
covariant derivative.}
\begin{equation}\label{comp1}
\Omega^{~~\dagger}_{~AB}~~=~~ -~ \Omega_{~BA}~~ .
\end{equation}
In this paper we will impose an additional reality condition on the
connection,
\begin{equation}\label{reality}
\Omega^{~~\dagger}_{~AB} ~=~ \Omega_{~AB}~~.
\end{equation}
Together with the metric compatibility condition (\ref{comp1}) the reality
condition implies the following conditions on the components $\Omega_{ABC}$
of the connection 1-form $\Omega_{AB}$ in the $E^A$ basis
\begin{eqnarray}\label{comp2}
\Omega_{~abc} & ~=~ & -~\Omega^{bac}, \cr
\Omega_{~ab {\dot 5}} & ~=~ & -~\Omega_{~ba{\dot 5}}~ = ~
\omega_{~ab {\dot 5}} e~~,\cr
\Omega_{~a{\dot 5}b} & ~=~ & - ~\Omega_{~{\dot 5}ab}, \cr
\Omega_{~a{\dot 5}{\dot 5}} & ~=~ & -~ \Omega_{~{\dot 5}a{\dot 5}} ~=~
\omega_{~a{\dot 5}{\dot 5}} e,\cr
\Omega_{~{\dot 5}{\dot 5}a}&~ =~ & \Omega_{~{\dot 5}{\dot 5}{\dot 5}} ~=~0 ~~.
\end{eqnarray}

That is to say, the reality condition (\ref{reality}) requires that the
internal indexed components of the connection 1-forms are ordinary functions.
\subsection{ The first structure equation and torsion 2-forms}

The first Cartan structure equation defines the torsion 2-forms $ T^A $ as
given by:
\begin{equation}\label{TORSION}
T^A = D E^A - E^B \wedge \Omega^A_{~~B}~,
\end{equation}
In Ref.\cite{VIWA}, we had assumed $ T^A = 0 ~~(A=a,{\dot 5})$ to determine the
connection $\Omega$. As noted before, we were lead to a theory with a tensor,
vector and scalar fields and also additional dilaton-like fields. In the
present paper, we shall assume
\begin{eqnarray}\label{TORFREE}
T_{abc} &~~=~~& T_{ab{\dot 5}} ~~=~~0  ~~,\cr
T_{{\dot 5}AB} &~~=~~& t_{{\dot 5}AB} r .
\end{eqnarray}

In other words, the torsion 2-forms involving the external physical spacetime
index vanish while the torsion 2-form involving the internal index
${\dot 5}$ as in Eq.(\ref{TORFREE}) does not vanish. Then we can determine
$ t^{\dot 5}_{~AB}$ as well as the hermitian and metric compatible connection
1-forms $ \Omega_{~AB}$ in terms of the vielbeins.

Using the general formula (\ref{DER1F}), it is straightforward to compute the
exterior derivatives $  DE^A$ needed to calculate $\Omega_{ABC}$ in
Eq.(\ref{TORSION}). We omit the details and give only the results.
\begin{eqnarray}\label{DERVIEL}
(DE_a)_{bc} &~~=~~& -~( DE_a)_{cb}~~=~~ {1\over 2}~\Big [~
(E^\mu_{~b}
E^\nu_{~c} - E^\mu_{~c}E^\nu_{~b})\partial_\mu E_{a\nu} ~~\cr
&~~~+~& m \Big(~( A_b
{\tilde E}^\nu_{~c} - A_c{\tilde E}^\nu_{~b})E_{a\nu} + (A_c\eta_{ab} -
A_b\eta_{ac})~\Big)~\Big ]~~,\cr
( DE_a)_{b{\dot 5}} &~~=~~& -~( DE_a)_{{\dot 5}b}~~=~~
{m\over 2} \Phi^{-1}( \eta_{ab} - {\tilde E}^\mu_b E_a^{~ \mu} )~~, \cr
( DE_{\dot 5})_{bc} &~~=~~& -~( DE_{\dot 5})_{cb}
{}~~=~~-~{1\over 2}~\Big [~(E^\mu_{~b}
E^\nu_{~c} - E^\mu_{~c}E^\nu_{~b})\Phi \partial_\mu A_\nu ~~\cr
&~~~+~& m (A_b
{\tilde E}^\nu_{~c} - A_c{\tilde E}^\nu_{~b})({\tilde A}_\nu - A_\nu )\Phi
{}~\Big]~~,\cr
( DE_{\dot 5})_{b{\dot 5}} &~~=~~& -~( DE_{\dot 5})_{{\dot 5}b}~~=~~
{1\over 2}~{\tilde E}^\mu_b \Big [~ {\partial_\mu \Phi \over \Phi } + m (A_\mu
-{\tilde A}_\mu \Phi {\tilde \Phi }^{-1})~ \Big ] ~~.
\end{eqnarray}

In component form, the first Cartan structure equation reduces to
\begin{equation}
T_{ABC} ~~=~~ (DE_A)_{BC} - {1\over 2 }( \Omega_{ABC} - \Omega_{ACB})~~.
\end{equation}
With the condition (\ref{TORFREE}) on the torsion 2-forms we obtain
\begin{equation}
\Omega_{abc} ~~=~~ ( DE_a)_{bc}+( DE_b)_{ca}- ( DE_c)_{ab}
{}~~,
\end {equation}
which in conjunction with Eqs.(\ref{DERVIEL}) determines $ \Omega_{abc}$ in
terms
of vielbeins.

The condition (\ref{TORFREE}) together with Eq.(\ref{comp2}) leads to
the following equation
\begin{equation}
\Omega_{ab{\dot 5}} - T_{{\dot 5} ab}~~ = ~~( DE_a)_{b{\dot 5}} +
( DE_b)_{{\dot 5}a}- ( DE_{\dot 5})_{ab}
{}~~,
\end{equation}
from which we can determine $ \Omega_{ab{\dot 5}} = \omega_{ab{\dot 5}}e $
and $ T_{{\dot 5}ab} = t_{{\dot 5}ab} r $ .

Finally, from
\begin{equation}
\Omega_{{\dot 5}c {\dot 5}} = 2 ( T_{{\dot 5} {\dot 5} c} - ( DE_{\dot 5}
)_{{\dot 5} c})
\end{equation}
we can determine $\Omega_{{\dot 5}c {\dot 5}}= \omega_{{\dot 5}c{\dot 5}} e $
and $T_{{\dot 5}{\dot 5} c} = t_{{\dot 5}{\dot 5} c} r$ .

The final results are as follows :

The components of the torsion 2-form $ T^{{\dot 5}}$ are
\begin{eqnarray}\label{TOR5}
T_{{\dot 5} ab} &~~=~~& {1\over 2}~ {\tilde E}^\mu_{~a}~\Big[~ {\tilde
E}^\nu_{~b} {\tilde \Phi} {\tilde F}_{\mu\nu} - E^\nu_{~b}\Phi F_{\mu \nu }
{}~\Big] + {m \over 4}~ \Big[ ~ (E^\mu_{~a}  {\tilde E}_{b\mu } - E^\mu_{~b}
{\tilde E}_{a\mu}) {\tilde \Phi}^{-1} \cr
&~~~+~& \Phi^{-1}( {\tilde E}^\mu_{~a\mu } - {\tilde E}^\mu_a E_{b\mu })
+ 2( {\tilde A}_\mu \Big(~(A_b  {\tilde E}^\mu_{~b} - A_b {\tilde E}^\mu_{~a})
\Phi - ( {\tilde A}_a E^\mu_{~b} - A_b E^\mu_{~a}) {\tilde \Phi}~\Big] ~~,\cr
T_{{\dot 5} a {\dot 5}} &~~=~~& {1\over 4} ~\Big [~ (~ {\tilde E}^\mu_{~b}{
\partial_\mu \Phi \over \Phi} - E^\mu_{~b}{\partial_\mu  {\tilde \Phi} \over
 {\tilde \Phi}}~) + m (~ {\tilde E}^\mu_{~b} A_\mu - E^\mu_{~b} {\tilde A}_\mu
\cr
&~~~+~& A_b {\tilde \Phi} \Phi^{-1} - {\tilde A}_b \Phi {\tilde
\Phi}^{-1}~)~\Big] ~~.
\end{eqnarray}

The components of the connection 1-forms $ \Omega_{AB} $ are given by:
\begin{eqnarray}\label{OMEGA}
\Omega_{abc} &~~=~~&  {1\over 2}~\Big[~E^\mu_{~b}E^\nu_{~c}
(\partial_\mu E_{a\nu}
-\partial_\nu E_{a\mu }) +  E^\mu_{~c}E^\nu_{~a} (\partial_\mu E_{b\nu}
-\partial_\nu E_{b\mu }) -  E^\mu_{~a}E^\nu_{~b} (\partial_\mu E_{c\nu}
-\partial_\nu E_{c\mu })~\Big] \cr
&~~~+~& {m \over 2}~ \Big[~ (A_b {\tilde E}^\nu_{~c} - A_c {\tilde E}^\nu_{~c}
)
E_{a\nu} +  (A_c {\tilde E}^\nu_{~a} - A_a {\tilde E}^\nu_{~c} )
E_{b\nu} -  (A_a {\tilde E}^\nu_{~b} - A_b {\tilde E}^\nu_{~a} )
E_{c\nu} \cr
&~~~+~& 2( A_a\eta_{cb} -A_b \eta_{ac})~ \Big] ~~,\cr
\Omega_{ab{\dot 5}}&~~=~~& {1\over 4}~ \Big (~E^\mu_{~a} E^\nu_{~b}
F_{\mu \nu} \Phi +
{\tilde E}^\mu_{~a}  {\tilde E}^\nu_{~b}
{\tilde F}_{\mu \nu}{\tilde \Phi}~\Big) +
{m \over 4}~ \Big[~ \Phi^{-1} ({\tilde E}^\mu_{~a} E_{b\mu} -
{\tilde E}^\mu_{~b} E_{a\mu}) \cr
&~~~+~& { \tilde \Phi}^{-1} ( E^\mu_{~a}{\tilde E}_{b\mu} -  E^\mu_{~b}
{\tilde E}_{a\mu}) + ({\tilde A}_\nu - A_\nu ) \Big(~( {\tilde A}_a E^\nu_{~b}
- {\tilde A}_b E^\nu_{~a}){\tilde \Phi} \cr
&~~~-~&   ( A_a  {\tilde E}^\nu_{~b}
-  A_b  {\tilde E}^\nu_{~a}) \Phi~ \Big)~ \Big ] \cr
\Omega_{{\dot 5}ab}&~~~=~& -~{1\over 4}~\Big(~E^\mu_{~a} E^\nu_{~b}
F_{\mu \nu} \Phi + {\tilde E}^\mu_{~a}  {\tilde E}^\nu_{~b}
{\tilde F}_{\mu \nu}{\tilde \Phi}~\Big)
+ {m \over 4}~ \Big[~ \Phi^{-1}\Big(~ ( 4\eta_{ab} - (3 {\tilde E}^\mu_{~b}
E_{a\mu} + {\tilde E}^\mu_{~a}
E_{b\mu})~\Big) \cr
&~~~+~& { \tilde \Phi}^{-1} ( E^\mu_{~b}{\tilde E}_{a\mu} +  E^\mu_{~a}
{\tilde E}_{b\mu}) - ({\tilde A}_\nu - A_\nu ) \Big(~ ( {\tilde A}_a E^\nu_{~b}
- {\tilde A}_b E^\nu_{~a}){\tilde \Phi} \cr
&~~~-~&   ( A_a  {\tilde E}^\nu_{~b}
-  A_b  {\tilde E}^\nu_{~a}) \Phi~ \Big)~ \Big] \cr
\Omega_{{\dot 5} b{\dot 5}} & ~~=~~& -~ \Omega_{b {\dot 5}{\dot 5}} ~~=~~
{1\over 2}~\Big[~{\tilde E}^\mu_b  {\partial_\mu \Phi \over \Phi }
+  E^\mu_b {\partial_\mu {\tilde \Phi} \over {\tilde \Phi}}~ \Big] +
{ m \over 2}\Big[~ {\tilde E}^\mu_b ( A_\mu
-{\tilde A}_\mu \Phi {\tilde \Phi }^{-1}) \cr
&~~~+~&  E^\mu_b ({\tilde A}_\mu
- A_\mu {\tilde \Phi} \Phi^{-1})~\Big ] ~~.
\end{eqnarray}
\subsection { Second structure equation, curvature and the action}
The second Cartan structure equation defines curvature 2-forms as follows
\begin{equation}\label{CARTAN2}
R_{AB} ~~=~~ D\Omega_{AB} + \Omega_{AC} \wedge \Omega^C_{~B}
\end{equation}
 It is straightforward to use the expressions for the connection 1-forms
given in
Eq.(\ref{OMEGA}) to compute the components $ R_{ABCD}$ of the curvature
2-forms. We recall from Ref.\cite{VIWA} the expression of the Ricci scalar
curvature \footnote{ The interested reader can see the Ref.\cite{VIWA} for
the expression of R in an inner product form.}
\begin{equation}
R~~=~~ \eta^{AC} R_{ABCD} \eta^{BD}.
\end{equation}
After a lengthy but straightforward calculation we obtain the final expression
of the generalized Ricci scalar curvature in the form
\begin{equation}\label{RICCI}
R ~~=~~ \pmatrix{ R_1 & 0 \cr
                   0 & R_2 \cr}~~=~~ R^{(0)} + R^{(1)} + R^{(2)},
\end{equation}
where $ R^{(0)},~R^{(1)},~R^{(2)} $ represent terms proportional to
$ m^0,~m,~m^2 $ respectively.
The explicit expressions of $ R^{(0)}, R^{(1)}$ and $ R^{(2)}$ are given as
follows:
\begin{eqnarray}\label{R0}
R^{(0)} &~~=~~&
{1\over 2}~\pmatrix{ r_1 & 0 \cr
                    0 & r_2 \cr} - {1\over 32}~\Big(~ 3 \Phi^2
F^2 +2 \Phi {\tilde \Phi}{\tilde E}^{a\mu}E^\rho_{~a}{\tilde E}^{b\nu }
E^\tau_{~b} {\tilde F}_{\mu \nu } F_{\rho \tau } -
{\tilde \Phi}^2 {\tilde F}^2~\Big)
\cr &~~~-~&
{1\over 2}~{\tilde E}^{a\mu} E^\nu_{~a} {\partial_\mu \Phi \over \Phi }
{\partial_\nu {\tilde \Phi} \over {\tilde \Phi}} +
{1\over 2} ~{\tilde E}^{a\mu} E^\nu_{~a} {\partial_\mu {\tilde \Phi} \over
{\tilde \Phi}} {\partial_\nu {\tilde \Phi} \over {\tilde \Phi}}
- {1\over 2}~{\tilde G}^{\mu \nu } {\partial_\mu \partial_\nu \Phi \over \Phi}
\cr &~~~-~&
{1\over 2}~{\tilde E}^{\mu a}E^\nu_{~a}{\partial_\mu \partial_\nu
{\tilde \Phi} \over {\tilde \Phi} } - {1\over 2}~ {\tilde E}^{a\mu}\partial_\mu
E^\nu_{~a} {\partial_\nu \Phi \over \Phi} - {1\over 2}~ {\tilde E}^{a\mu}
\partial_\mu E^\nu_{~a} {\partial_\nu {\tilde \Phi} \over {\tilde \Phi}} \cr
&~~~+~& {1\over 4}~\Big(~{\tilde E}^\mu_{~a}{\partial_\mu \Phi \over \Phi}
+ {\tilde E}^\mu_{~a}{\partial_\mu {\tilde \Phi} \over {\tilde \Phi}}~\Big)
\Big[~E^{a\nu} E^{b\rho}( \partial_\rho E_{b\nu} - \partial_\nu E_{b\rho}) \cr
&~~~+~& {\tilde E}^{a\nu} {\tilde E}^{b\rho}( \partial_\rho {\tilde E}_{b\nu} -
\partial_\nu {\tilde E}_{b\rho})~\Big],
\end{eqnarray}

\begin{eqnarray}\label{R1}
R^{(1)} &~~=~~&
m \bigg [~{1\over4}~ (\partial_\mu E_{a\nu} - \partial_\nu E_{a\mu})
\Big(~
3 E^{a\nu} E^\mu_{~b}{\tilde E}^{b\rho}A_\rho - 4 E^{a\nu} A^\mu
({\tilde E}^{b\rho}E_{b\rho}) - 4 E^\mu_{~b}A^a {\tilde E}^{b\nu }
\cr &~~~+~&
4 A^\mu E^a_{~\rho}{\tilde E}^{b\rho} E^\nu_{~b} + 8 A^\mu
{\tilde E}^{a \nu } + 12 A^\nu E^{a\mu } +
E^{a\mu}G^{\nu \rho}{\tilde A}_\rho
- E^{b\nu} E^{a\mu}{\tilde A}_b \Phi{\tilde \Phi}^{-1}
\cr  &~~~-~&
 A^\nu E^{a\mu }{\tilde \Phi} \Phi^{-1}~\Big)
+{1\over 4}~ (\partial_\mu {\tilde E}_{a\nu} - \partial_\nu {\tilde E}_{a\mu})
\Big(~4 {\tilde E}^{a\mu} A^b {\tilde E}^{\nu }_{~b} +
{\tilde E}^{a\mu}{\tilde G}^{\nu \rho} A_\rho
\cr &~~~-~&
{\tilde A}^\nu {\tilde E}^{a\mu } \Phi {\tilde \Phi }^{-1}
+  {\tilde E}^{a\mu}
{\tilde E}^{b\nu} E^{\rho}_{~b}{\tilde A}_\rho
- {\tilde E}^{b\nu }{\tilde E}^{a\mu} A_b
\Phi^{-1}{\tilde \Phi}~\Big) - {1\over 4} F_{\mu\nu}~\Big(~{5\over 2}~
{\tilde E}^{a\mu} E^\nu_{~a}
\cr &~~~+~&
{3\over 2}~\Phi {\tilde \Phi}^{-1} G^{\nu\rho}
E^{b\mu }{\tilde E}_{b\rho }
+ ( {\tilde A}_\rho - A_\rho)~( 2\Phi  {\tilde \Phi} G^{\mu\rho}
{\tilde A}_a E^{a\nu} + 3 \Phi^2 A^\mu E^{a\nu} {\tilde E}^\rho_{~a} )~\Big)
\cr &~~~+~&
{1\over 4}~ {\tilde F}_{\mu \nu}~\Big(~{3\over 2} ( {\tilde A}_\rho -A_\rho)
\Phi {\tilde \Phi}
A^b {\tilde G}^{\mu \rho}  {\tilde E}^\nu_{~b}
- {1\over 2}~ {\tilde \Phi} \Phi^{-1} {\tilde G}^{\mu \rho}
{\tilde E}^{a\nu }E_{a\rho} + {1\over 2} {\tilde E}^{a\mu} E^\nu_{~a}
\cr &~~~+~&
{\tilde \Phi}^2 ({\tilde A}_\rho - A_\rho)
{\tilde A}^\mu {\tilde E}^{b\nu}E^\rho_b ~\Big)
+{1\over 2}~{\tilde E}^{a\mu }\partial_\mu \Big( {\tilde E}^{\rho}_{~a}
(A_\rho - {\tilde A}_\rho \Phi {\tilde \Phi}^{-1}) + E^\rho_{~a}
( {\tilde A}_\rho - A_\rho {\tilde \Phi}\Phi^{-1})~\Big)
\cr  &~~~-~&
E^\mu_{~a} \partial_\mu \Big(~ A_\rho {\tilde E}^{a\rho} - A^a (
{\tilde  E}^{b\rho} E_{b\rho}) + 3 A^a~\Big )
+{1\over 4}~ {\partial_\mu \Phi \over \Phi}~\Big( ~3A^a  {\tilde E}^\mu_{~a}
- A^a  {\tilde E}^\mu_{~a}({\tilde E}^{b\rho}E_{b\rho})
\cr &~~~+~&
 3 {\tilde E}^{a\mu}E^\rho_{~a} {\tilde A}_\rho
- {\tilde A}^\mu (E^{b\rho} {\tilde E}_{b\rho}) + 3 {\tilde A}^\mu
+ {\tilde G}^{\mu\rho} A_\rho - 2  {\tilde E}^{a\mu} A_a  {\tilde \Phi}
\Phi^{-1}~\Big) \cr
&~~~+~& {1\over 4}{\partial_\mu {\tilde \Phi}\over  {\tilde \Phi}}
{}~ \Big(~ 3A^\mu -A^\mu ({\tilde E}^{b\rho}E_{b\rho})
+ G^{\mu\rho}  {\tilde A}_\rho -  {\tilde A}^a E^{b\rho} {\tilde
E}_{b\rho}E^\mu_{~a} + 3 {\tilde A}^a E^\mu_{~a} \cr
&~~~-~&  {\tilde
E}^{a\rho}E^\mu_{~a}A_\rho
+ 2\Phi {\tilde \Phi}^{-1} E^\mu_{~a}  {\tilde A}^a ~\Big)~\bigg]~~ ,
\end{eqnarray}

\begin{eqnarray}\label{R2}
R^{(2)}
&~~=~~&
{m^2 \over 16} ~\bigg[~ \Phi^{-2}~\big(- 32 +
48 {\tilde E}^{b\rho} E_{b\rho}
- 7{\tilde E}^{b\mu} E_{a\mu} {\tilde E}^{a\rho} E_{b\rho} - {\tilde G}^{\mu
\nu} G_{\mu \nu } - 8 ({\tilde E}^{b\rho} E_{b\rho} )^2~\big)
\cr &~~~+~&
{\tilde \Phi}^{-2}~\big( E^{b\rho}{\tilde E}_{a\rho} E^{a\nu} {\tilde
E}_{b\nu} - G^{\mu \nu } {\tilde G}_{\mu \nu } \big) - 2 \Phi^{-1}{\tilde
\Phi}^{-1}~\big( 4 - {\tilde E}^{a\rho} E_{b \rho} {\tilde E}^b_{~\nu}
E^\nu_{~a}\big)~\bigg]
\cr &~~~+~&
{m^2\over 4} ( {\tilde A}_\mu - A_\mu) \bigg[~ A_\rho
{\tilde G}^{\mu\rho} -
A_a {\tilde E}^{a\rho} E_{b\rho}{\tilde E}^{b\mu } + \Phi {\tilde \Phi}^{-1}
\big( E^\nu_{~a}{\tilde E}^{a\mu} A_b{\tilde E}^b_{~\nu} - A^\mu \big)~\bigg]
\cr &~~~+~&
{m^2 \over 4}~\bigg[~ - 6 A_\nu {\tilde E}^{b\nu}E_{b\rho}{\tilde E}^{a
\rho} A_a - A_\nu {\tilde E}^{a\nu}A_a - 2 A^2 ({\tilde E}^{b\rho}E_{b\rho})^2
+ 4 A^2 ({\tilde E}^{b\rho}E_{b\rho})
\cr &~~~-~& 12 A^2 + 3 A_\nu A_a {\tilde E}^{a\nu}({\tilde E}^{b\rho}E_{b\rho})
+ 3 A^2 {\tilde E}^{b\nu}E_{b\mu }{\tilde E}^{a\mu}E_{a\nu} -
3 A_a{\tilde E}^{a\rho} A_b {\tilde E}^{b\nu} G_{\rho \nu}
\cr &~~~+~&
3 A^2{\tilde G}^{\mu\nu}G_{\mu\nu}
+ 7{\tilde A}_\rho A^\rho - 4 {\tilde A}_a A^a ({\tilde E}_{b\rho}E^{b\rho})
+ 10 {\tilde A}^a A_a - {\tilde A}^\mu A_\mu ({\tilde E}_{b\rho}E^{b\rho})
\cr &~~~-~&
 A^\mu {\tilde A}_\mu({\tilde E}^{b\rho}E_{b\rho})
+ 3 {\tilde A}^\mu A_\mu
+  {\tilde A}_\nu G^{\mu \nu } {\tilde A}_\mu - {\tilde A}_\mu  E^{a\mu}
 {\tilde A}_a ({\tilde E}_{b\rho}E^{b\rho})
\cr &~~~+~&
3 {\tilde A}_\rho E^{a\rho}{\tilde A}_a
+4 A_\mu{\tilde A}^\mu \Phi {\tilde \Phi}^{-1} -2 {\tilde A}^2 \Phi^2 {\tilde
\Phi}^{-2} + 2 {\tilde A}_a E^{a\rho}{\tilde A}_\rho \Phi {\tilde \Phi}^{-1}
\cr &~~~+~&
2 A_a {\tilde E}^{a\mu} A_\mu {\tilde\Phi} \Phi^{-1}~\bigg]
- {m^2 \over 4}~\big ({\tilde A}_a\Phi {\tilde \Phi }^{-1} + A_a {\tilde \Phi}
\Phi^{-1}\big)~\bigg[~ A_\nu{\tilde E}^{a\nu} - A^a
({\tilde E}^{b\rho}E_{b\rho})
\cr &~~~+~&
3 A^a + {\tilde A}_\nu E^{a\nu} -
{\tilde A}^a ({\tilde E}_{b\rho}E^{b\rho}) + 3 {\tilde A}^a ~\bigg]~
\cr &~~~+~&
{m^2 \over 16} \big({\tilde A}_\mu - A_\mu \big) ({\tilde A}_\nu - A_\nu)~
\bigg[~ 6\Phi {\tilde \Phi}\big( A^\mu {\tilde A}^\nu - {\tilde A}^a A_a
{\tilde E}^{a\mu} E^\nu_{~a} \big)
\cr  &~~~+~&
5 \Phi^2 ~\big( A^2 {\tilde G}^{\mu \nu} - A_b {\tilde E}^{b\mu } {\tilde E}^{a
\nu} A_a) + {\tilde \Phi}^2~\big( {\tilde A}^2 G^{\mu \nu} - {\tilde A}_a
{\tilde A}_b E^{a\nu} E^{b\mu }\big)~\bigg]~~,
\end{eqnarray}
where
\begin{equation}
F_{\mu \nu }= \partial_\mu A_\nu - \partial_\nu A_\mu = \pmatrix{ f_{1\mu\nu} &
0 \cr
0 & f_{2\mu\nu} \cr} ~~,
\end{equation}
and $r_1$ and $ r_2$ are the ordinary Ricci scalar curvatures on the first and
second copies of spacetime, respectively.

The volume element is given by
\begin{equation}\label{volume}
D^5X~=~D^4X \sqrt{-det | {\cal G}|}
\end{equation}
Here $det|{\cal G}|$ denotes the determinant of our generalized metric
defined in Eq.(\ref{METRIC}) and is given by
\begin{eqnarray}
det|{\cal G}|& \doteq &{1\over 5!}{\epsilon }_{N_1 N_2 N_3 N_4 N_5}
{\epsilon}_{M_1 M_2 M_3 M_4 M_5}{\cal G}^{N_1 M_1}{\cal G}^{N_2 M_2}
{\cal G}^{N_3 M_3} {\cal G}^{N_4 M_4} {\cal G}^{N_5 M_5} \cr
      & = &
{1\over 4!}{\epsilon}_{ \nu_1 \nu_2 \nu_3 \nu_4}
{\epsilon}_{\mu_1
\mu_2 \mu_3 \mu_4 } {\cal G}^{\nu_1\mu_1} {\cal G}^{\nu_2 \mu_2}
{\cal G}^{\nu_3 \mu_3}{\cal G}^{\nu_4
\mu_4} {\cal G}^{55} \equiv  det|G|\Phi~{\bf 1},
\end{eqnarray}
where ${\epsilon}$'s are the fully antisymmetric Levi-Civita tensors and
\begin{equation}
det|G| ~~=~~ \pmatrix{ det |g_1| & 0 \cr
                          0 & det |g_2| \cr }.
\end{equation}

The action then is defined as
\begin{eqnarray}\label{action}
S &~~=~~& {1\over m. \kappa}~Tr~(\int dx^4 \sqrt{-det~G}~ R)~~,\cr
  &~~=~~&  S_1 + S_2 ~~,\cr
S_1&~~=~~& \sqrt{-det |g_1|}\varphi_1 R_1~~,\cr
S_2&~~=~~& \sqrt{-det |g_2|}\varphi_2 R_2~~,
\end{eqnarray}
where $\kappa = 16\pi^2G^{-2}/m $ and $ G $ is the Newton
constant.

The integration over the discrete space follows naturally to be
${1\over m} Tr $.

\setcounter{equation}{0}
\section{ Mass terms:}
The full action of our model (\ref{action}) contain six independent interacting
fields $ e^a_{1\mu},~e^a_{2\mu},~a_{1\mu},~a_{2\mu},
{}~\varphi_1 $ and $ \varphi_2 $.
Since the full expression for the Ricci scalar curvature $ R $ in
Eqs.(\ref{RICCI})-(\ref{R2}) is obviously extremely complex, here we will
concentrate on the massive modes in our model. We will concentrate on the
gravity sector first.
\subsection{ Gravity and massive tensor field}
To find the mass content of the tensor field we consider the part of the
action that contains only tensor fields. It turns out to be
\begin{eqnarray}\label{TENSORS}
R_t &~~=~~& \int dx^4~\sqrt{-det |G|}~ \bigg[ ~{1\over 2}~
\pmatrix{ r_1 & 0 \cr
           0 & r_2 \cr} +
{m^2 \over 16} ~\bigg(~ - 40 +
48 {\tilde E}^{b\rho} E_{b\rho}
\cr &~~-~~& 7{\tilde E}^{b\mu} E_{a\mu} {\tilde E}^{a\rho} E_{b\rho} -
{\tilde G}^{\mu
\nu} G_{\mu \nu } - 8 ({\tilde E}^{b\rho} E_{b\rho} )^2~
\cr &~~~+~&
 E^{b\rho}{\tilde E}_{a\rho} E^{a\nu} {\tilde
E}_{b\nu} - G^{\mu \nu } {\tilde G}_{\mu \nu }
+ 2 {\tilde E}^{a\rho} E_{b \rho} {\tilde E}^b_{~\nu}
E^\nu_{~a}~\bigg)~\bigg].
\end{eqnarray}

{}From the terms proportional to $ m^2$, we can see that $ e^\mu_{1a} $
and $ e^\mu_{2a}$ are not the fields corresponding to mass eigenstates since
their products appear in these terms giving rise to mixing.
To find mass eigenstates we write
\begin{eqnarray}\label{EPM}
E^\mu_{~a}&~~=~~& {1\over 2}~ \bigg(~e^\mu_{+a} {\bf 1} + e^\mu_{-a}
r~~\bigg)~~,\cr
{\tilde E}^\mu_{~a}&~~=~~& {1\over 2}~\bigg(~ e^\mu_{+a} {\bf 1} +
e^\mu_{-a} r~~\bigg)~~,
\end{eqnarray}
and substitute for them in Eq.(\ref{TENSORS}). We note that a proper mass term
has the general form   $ m^2 b^{a\mu} b_{a\mu} $, where $ b^a_{~\mu}$
represents the massive tensor field.

With this in mind, we find two possibilities for identifying the massive
fields:

i) If we choose $ e^\mu_{+a} $ as the vielbein for the metric that represents
gravity, we find the mass term for the tensor field $e^a_{-\mu}$ as
$ \sim 15/16 m^2 e^{a\mu}_- e_{-a\mu}$ in Eq.(\ref{TENSORS}). The terms in pure
$ e^a_{+\mu} $ give a cosmological constant. In the case we are considering,
these terms and the constant term cancel and consequently there is no
cosmological constant. Further
we note that, in the vacuum $ e_+$ is a physical field as $ e_- \rightarrow 0 $
and $ e^\mu_{+a} \rightarrow \delta^\mu_a $.

ii) If we choose $ e^\mu_{-a} $ as the vielbein for the gravity metric. The
same terms that give a mass to $ e^\mu_{-a}$ in the previous case, now becomes
the mass terms for $ e^a_{+\mu}$. Since the terms in pure $ e^a_{+\mu}$ and the
constant terms do not cancel, there is a cosmological constant in this case.
In vacuum, $ e^a_{+\mu}\rightarrow 0 $ and $e^\mu_{-a} \rightarrow
\delta^\mu_{~a} $.
The mass term for $ e^\mu_{+a}$ in this case is $ -9/16 m^2 e^{a\mu}_+
e_{+a\mu} $. There are also quartic terms in $ e^\mu_{+a}$. It would be
interesting to see whether this negative mass terms lead to spontaneous
symmetry breaking patters.

In the two limiting cases, when the massive tensor field is set to zero we
have the usual Einstein theory with the vielbein $e^\mu_{-a}$ or the
theory with the vielbein $e^\mu_{+a}$ together with a cosmological constant.

Now we will consider the mass terms of the vector and scalar fields with
the above two choices.

\subsection{ Mass terms of vector and scalar fields}

At classical level the tensor fields do not alter the mass terms of
vector and scalar fields. Hence we will turn off the tensor fields and consider
two limiting cases $ E^\mu_{~a} = \delta^\mu_{~a}$ and
$ E^\mu_{~a}= \delta^\mu_{~a}r $.
After inserting the particular $ E^\mu_{~a}$ into the the expression for
$R^{(2)}$ we find:

i) $ E^\mu_{~a}=\delta^\mu_{~a} $ : There is no mass terms for the scalar
fields. However, the mass term for vector fields is $ 4 m^2 a_{-\mu}^2 $ where
$ a_{\pm \mu} = 1/2 ( a_{1\mu}-a_{2\mu})$. This means that in this case
$ a_{+\mu}$ is massless and $ a_{-\mu}$ is massive.

ii) $ E^\mu_{~a} =\delta^\mu_{~a}r$ : The mass terms in this case are given
by
\begin{equation}
R^{(2)}~~=~~ -96m^2 \varphi_{-}^{2} - 36 m^2 a_{-}^2~~,
\end{equation}
where $ \varphi_{\pm} = \varphi_1 \pm \varphi_2 $.

The action for this part is
\begin{equation}
S_m ~~\sim~~  \int dx^4  -96 m^2(\varphi_{-}^{3} +\varphi_{+}\varphi_{-}^{2})
- 36 m^2( \varphi_{+} + \varphi_{-}) a_{-}^2 ~~.
\end{equation}
Note that in vacuum $ \varphi_+ = 1 $. Therefore  $ \varphi_{-} $ is the
physical mode while we have to expand $\varphi_{+}$ in terms of the physical
field $\sigma $ as follows:
\begin{equation}
\varphi_{+} ~~=~~2 exp(-\sigma ) ~~,
\end{equation}
where in vacuum $\sigma \rightarrow 0 $.

Using this expansion, the mass terms of $ a_{-\mu}$ and $ \varphi_{-}$
are $-36 m^2 a^2_{-}$ and $ -96m^2 \varphi^2_{-}$ respectively. These mass
terms
as well as the mass term for the tensor field $e^\mu_{+a}$
in this case are negative. It would be interesting to include the quartic terms
to see whether these negative mass terms lead to some spontaneous symmetry
breaking patterns. The quartic potential for vector fields are already there
in Eq.(\ref{R2}). To have the quartic potential for the scalar field, however,
one has to modify the wedge product of forms in Eq.(\ref{wedge}). Such
modifications will be discussed elsewhere.

\setcounter{equation}{0}
\section{Summary and Conclusions:}

We have in the  previous papers \cite{VIWA,LVW} developed a discretized
version of
Kaluza-Klein theory by replacing the continuous fifth dimension by two discrete
points. In the language of NCG, we may speak of two copies of spacetime instead
of an infinite number of them in the standard Kaluza-Klein theory ( For every
internal point in the fifth dimension we have a four-dimensional spacetime).
The geometry of the extended spacetime permitted us to introduce a generalized
vielbein consisting of a pair of tensor, a pair of vector and a pair of scalar
fields. When we imposed the standard metric compatibility and torsion free
conditions to determine the connection 1-form, we found constraints on the
vielbeins in the form of dynamical dilaton fields that implied new and
interesting consequences.

In the present paper we have pursued the investigation further to see whether
we can eliminate the constraints on the vielbeins by relaxing the torsion free
condition. In order to remain as close to the Riemannian geometry as possible,
we still require that the torsion 2-forms corresponding to the physical
spacetime do vanish. However, by making an ansatz about torsion 2-form
corresponding to the internal space, we determine uniquely not only all the
connection 1-form coefficients, but also the nonvanishing torsion components in
terms of the assumed vielbeins. This is in contrast to the usual Riemannian
geometry where nonvanishing torsion does not lead to a unique determination of
the connection coefficients.

With the unique determination of the connection coefficients, we obtain a
Lagrangian and an action that has a rich and complex structure with interacting
tensor, vector and scalar fields. It appears as sum of two terms $ S_1$ and $
S_2$, each consisting of all the six independent fields and each representing a
generally covariant action. In $ S_1 (S_2)$, the vierbein $e^\mu_{1a}~(
e^\mu_{2a})$ acts as the metric field with appropriate kinetic term while the
other $e^\mu_{2a}~(e^\mu_{1a}) $ coupled to $e^\mu_{1a}~(e^\mu_{2a})$ in
quadratic and quartic terms. This suggests that $ e^\mu_{1a}$ and
$e^\mu_{2a}$ are not eigenstates of mass. Instead we have two mass eigenstates
as $e^\mu_{\pm a}= e^\mu_{1a} \pm e^\mu_{2a} $. We have two possibilities
of choosing $e^\mu_{+a} $ or $ e^\mu_{-a}$ as representing the gravity field.
In the first case, $e^a_{-\mu}$ and $ a_{-\mu} $ are massive fields
while the scalar fields and $a_{+\mu} $ are massless. There is no cosmological
constant in this case. In the second case, there is a cosmological
constant and negative mass terms for tensor, vector and scalar fields.

In conclusion, we like to observe that our discretized version of Kaluza-Klein
theory within the framework of NCG demonstrates an extremely promising approach
to internal structure of elementary particles. If the internal space is
discrete, one obtains only a finite number of massive modes and thus avoids the
problem of infinite number of massive modes and of the necessity of truncation.
In addition to having mass, the fields have interactions proportional to the
mass parameter $ m $ and the Newton constant $ G$.It is extremely interesting
to explore the consequences of such theory on gravity. The highly correlated
interactions also suggest strong quantum implications that are fascinating to
study.

\noindent
{\bf Acknowledgments.}

This work was supported in part by the U.S. Department of Energy under contract
number DE-FG02-85ER40231. One of the authors ( N.A.V.) thanks Profs T.N.Truong
and Y.X. Pham for financial support and invitation to give talks
on NCG at \'Ecole Polytechnique and University Paris VI where the hospitality
and stimulating discussions inspired many ideas in this paper. K.C.W.
would like to thank the Fulbright Foundation for a grant and Dr.G.C.Joshi for
many useful discussions and his hospitality at the University of Melbourne,
Australia where this work was partially done. The contribution of G.Landi at
the
initial stage of this research program, mainly in Ref.\cite{LVW} is greatly
appreciated.

\end{document}